# Magnetic and electronic phase transformations in $(Sm_{0.65}Sr_{0.35})MnO_3$ induced by temperature and magnetic field

E. M. Levin,[1,2*] P. M. Shand [3]

[1] Department of Physics and Astronomy, [2] Ames Laboratory DOE, Iowa State University, Ames, IA, 50011-3020, USA
[3] Department of Physics, University of Northern Iowa, Cedar Falls, IA 50614-0150, USA

Temperature (4.2-260 K) and magnetic field (0-50 kOe) dependencies of the dc electrical resistance, dc magnetization, and ac magnetic susceptibility of $(Sm_{0.65}Sr_{0.35})MnO_3$ prepared from high purity components have been studied. $(Sm_{0.65}Sr_{0.35})MnO_3$ undergoes a temperature induced transition between ferromagnetic metallic and paramagnetic insulating-like states. A magnetic field strongly affects this transition resulting metallic state and "colossal" magnetoresistance in the vicinity of the metal $\leftrightarrow$ insulator transformation. Magnetic and electric properties of $(Sm_{0.65}Sr_{0.35})MnO_3$ are different compared to those reported earlier for similar composition, which is probably attributable to the purity of the starting materials and/or different process of annealing. The character of phase transformations observed in $(Sm_{0.65}Sr_{0.35})MnO_3$ is compared to that reported for $Gd_5(Si_xGe_{4-x})$ intermetallic alloys with a true first order phase transition.

PACS numbers: 72.20.My, 72.80.-r, 75.30.Kz

* - corresponding author, E-mail: levin@iastate.edu

## I. INRODUCTION

During last decade significant interest has been focused on the complex manganese oxides $(R_{1-x}A_x)MnO_3$ series where R is a trivalent rare-earth cation and A is a divalent, alkaline-earth cation.[1] Initially this interest was sparked by the discovery of large negative magnetoresistance, so-called colossal magnetoresistance (CMR).[2] However, later it became clear that there are many additional features, including a first order phase transition (FOPT), which attracts more attention to that class of materials. First, the electronic phase transition between the insulating and metallic states can be induced by temperature and/or magnetic field. Second, the electronic phase transition can be accompanied by a magnetic order $\leftrightarrow$ disorder transition and parameters of both transitions are very sensitive to the oxide's composition. Third, the members of the $(R_{1-x}A_x)MnO_3$ series have a distinctly layered crystal structure that forms a two-dimensional magnetic system, which should play an important role in phase transformations. Finally, manganese oxides allow substitution for Mn and 4f-paramagnetic elements by various other elements, which makes them a unique model system for systematic studies of FOPT in materials with a layered crystal structure.

Similar behavior related to first order phase transitions can be found in some metallic materials with a layered crystal structure where 3d- and 4f-paramagnetic ions form a two-dimensional magnetic system.[3,4] The complexity of such materials typically is reflected by a magnetic phase diagram showing their magnetic state vs. temperature, magnetic field, pressure, and composition. Magnetic phase diagrams of manganese oxides are, in general, more complex than those of metallic materials (see, for example, magnetic diagrams of $R_{1-x}A_xMnO_3$ series with R = Pr, Sm, and A = Ca, Sr reported by Martin et al.[5]). However, lanthanide-based intermetallic compounds with FOPT also may have non-trivial magnetic diagram[6-8] reflecting, therefore, some link between the nature of FOPT observed in these quite different materials. Hence, the analysis of both the manganese oxides and complex intermetallic compounds may aid in understanding materials where magnetic, electronic, and structural phase transformations can be induced by a magnetic field and temperature. That conclusion can be confirmed by recently reported data on instability of phase transformations against thermal cycling in quite different materials: $Pr_{0.5}Ca_{0.5}Mn_{1-x}Cr_xO_3$ manganese oxide (Mahendiran et al.,[9] see also the review by Raveau et al.[10]) and $Gd_5(Si_{1.95}Ge_{2.05})$ metallic alloy (Levin et al.[11]). In addition, similar martensitic-like field-induced transformations from the antiferromagnetic to ferromagnetic phase were recently discussed by Hardy et al.[12] for the $Pr_{0.6}Ca_{0.4}Mn_{0.96}Ga_{0.04}O_3$ and $Gd_5Ge_4$ compounds. Note that $Gd_5Ge_4$ is one of the representatives of the $Gd_5(Si_xGe_{4-x})$ family of intermetallic compounds where several interesting thermal, magnetic, and magnetoelectronic phenomena have been observed due to a first order phase transition.[7,11-19]

Manganese oxides have a quite different electronic structure than metallic alloys but in both types of materials one can observe similar peculiarities. In a certain range of the R ↔ A substitution in $(R_{1-x}A_x)MnO_3$, the transition between the ferromagnetic metal and paramagnetic insulator occurs, showing strong correlations between magnetic and electronic properties. Large negative magnetoresistance of manganese oxides observed in the vicinity of the magnetic and electronic phase transformations is a result of different electrical resistance of the initial insulating phase and the transformed metallic phase. A similar mechanism for magnetoresistance takes place in $Gd_5(Si_xGe_{4-x})$ intermetallic alloys where a magnetic field applied just above $T_C$ simultaneously with the paramagnet → ferromagnet transition associated with a FOPT affects the transformation from high- to low- or from low- to high-resistance states observed for $Gd_5(Si_2Ge_2)$[7] or $Gd_5(Si_{1.5}Ge_{2.5})$,[15] respectively. Note that the characteristics of FOPT even in metallic materials could strongly depend on the purity of used components as was suggested for $Gd_5Ge_4$ (compare the data reported in Refs. 17 and 20).

Among well-known manganese oxides, the $(Sm_{1-x}Sr_x)MnO_3$ series corresponds to a large value of the A-site cationic radius $\langle r_A \rangle$ which increases from 0.1132 nm for $x = 0$ to 0.1310 nm for $x = 1$. These materials exhibit CMR mainly on the hole doped side for $x$ varying from 0.20 to 0.52.[5] The ground magnetic state of $(Sm_{1-x}Sr_x)MnO_3$ and the character of both the electronic and magnetic phase transformations induced by temperature and/or magnetic field strongly depend on both the composition and distortion of the crystal lattice. The magnetic phase diagram of $(Sm_{1-x}Sr_x)MnO_3$ shows that samples with $0.30 < x < 0.52$ have a ferromagnetic metallic (FMM) ground state while sample with $x < 0.30$ samples have a ferromagnetic insulating (FMI) ground state. The occurring of the FMM state in $(Sm_{1-x}Sr_x)MnO_3$ is explained by the increase of $\langle r_A \rangle$ and, consequently, of both the Mn-O-Mn angle and bandwidth.[5]

Earlier, Thomas *et al.*[21] showed that $(Sm_{0.65}Sr_{0.35})MnO_3$ exhibits a metal ↔ insulator transition at 63 K and the Curie temperature, $T_C$, is 85 K. Later, Borges *et al.*[22] reported that $(Sm_{0.65}Sr_{0.35})MnO_3$ shows a metal ↔ insulator transition at temperature around 100 K and $T_C$ of 120 K indicating that a purity of used components and/or technology of the sample synthesis can be critical. It was also stated that neither a purely activated law, $R \sim R_0 \exp(\Delta\varepsilon/kT)$, nor a simple hopping law, $R \sim R_\infty \exp(T_0/T)$, are able to fit well the behavior of the electrical resistance in the insulating phase. Hence, in addition to the role of purity and technology, it should be interesting to study how a magnetic field will affect the electrical resistance and magnetization of $(Sm_{0.65}Sr_{0.35})MnO_3$ prepared from high quality components. Here, we report data on the crystal structure (at 293 K), temperature (4.2-260 K) and dc magnetic field (up to 50 kOe) dependencies of the electrical resistance and magnetization, and temperature dependencies of the real and imaginary components of the ac magnetic susceptibility of $(Sm_{0.65}Sr_{0.35})MnO_3$. The phase transformations induced by temperature and magnetic field in $(Sm_{0.65}Sr_{0.35})MnO_3$ are compared to those reported earlier for various manganese oxides and for $Gd_5(Si_xGe_{4-x})$ intermetallic alloys with a true first order phase transition.

## II. EXPERIMENTAL DETAILS

The $(Sm_{0.65}Sr_{0.35})MnO_3$ sample was prepared from 4.914 g of $Sm_2O_3$ (MPC, 99.999%), 2.241 g of $SrCO_3$ (Aldrich, 99.99%) and 3.774 g of $MnO_2$ (Aldrich, 99.99%) ball-milled together for 2 hours. After ball milling, the powder was pressed at pressure of $4\times10^8$ Pa into a pellet and annealed in air at 950°C for 21.5 hours. The obtained material was ground and passed through a 40-µm screen, then pressed again at a pressure of $4\times10^8$ Pa into pellets, and annealed at 1250°C for 6 days. X-ray powder diffraction analysis of the annealed sample was carried out on the $Cu_{K\alpha}$ line with $\lambda = 0.154178$ nm. Figure 1 shows the x-ray powder diffraction pattern of $(Sm_{0.65}Sr_{0.35})MnO_3$ collected at 300 K.

The material is single phase within the sensitivity of the technique, 1 to 2 vol. % of an impurity phase. The compound belongs to space group P2/c with the following unit cell parameters determined as a result of Rietveld refinement: $a = 0.54381(3)$, $b = 0.54485(3)$, $c = 0.76773(3)$ nm, and $\beta = 89.99(2)°$. This result is different from that reported by Borges *et al.*[22] where $(Sm_{0.65}Sr_{0.35})MnO_3$ was indexed as a cubic perovskite-type structure with the unit cell dimension $a = 0.5428$ nm. It is feasible that the monoclinic distortion of the ideal cubic lattice observed here is the result of either or both different purity of starting components as well as different heat treatment regime.

The sample for the electrical measurements had dimensions of approximately $1\times3\times6$ mm$^3$. The electrical connections to the sample were made by attaching thin platinum wires with H20E EPOTEK silver paste manufactured by Epoxy Technology. The dc electrical resistance measurements were carried out using Lake Shore Model 7225 magnetometer equipped with a probe for making four-point measurements.

The measurements were performed at constant dc electrical current of 10 mA in the temperature range from 4.2 to 260 K and in magnetic fields from 0 to 40 kOe with the current applied in opposite directions to eliminate possible thermal effects. The magnetic field vector was oriented parallel to the direction of electrical current. Samples for the magnetization and magnetic susceptibility measurements weighed ~0.2 g. Based on the shape of the sample we estimated the demagnetization factor to be ~0.3.

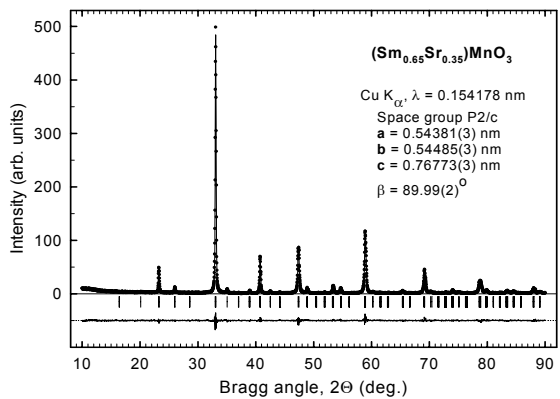

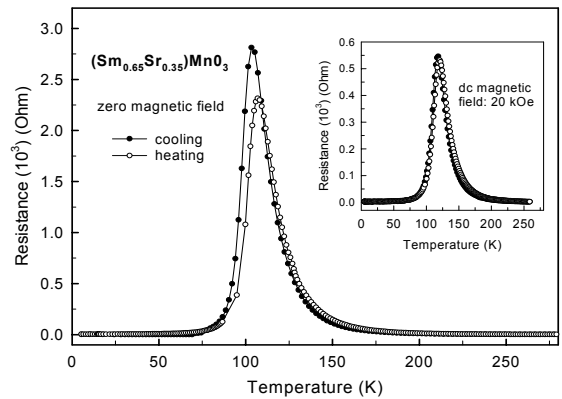

FIG. 1. Room temperature x-ray powder diffraction patterns of $(Sm_{0.65}Sr_{0.35})MnO_3$. The points represent observed data and the line drawn through the data points corresponds to the calculated pattern. The difference between the observed and calculated intensities is shown as the line at the bottom of the plot in the same scale as the observed data. The vertical bars located under the diffraction pattern indicate calculated positions of the $K_{\alpha 1}$ components of Bragg peaks.

FIG. 2. Temperature dependencies of the electrical resistance of $(Sm_{0.65}Sr_{0.35})MnO_3$ measured on cooling and heating in zero magnetic field. The inset shows temperature dependence of the electrical resistance measured in the 20 kOe magnetic field.

Measurements of the magnetization, dc and ac magnetic susceptibility were carried out using a Lake Shore magnetometer (magnetic field can be changed from 0 to 50 kOe). The rate of temperature change was ~1.5 K/min for both cooling and heating in the temperature range between 4.2 and 260 K. The ac magnetic susceptibility was measured in 12.5 and 25 Oe ac magnetic fields, at 125 Hz. The errors of the electrical and magnetic measurements were about 1 %.[7,17]

## III. EXPERIMENTAL RESULTS AND DISCUSSION

### A. Electrical resistance in magnetic field

Temperature dependencies of the electrical resistance of $(Sm_{0.65}Sr_{0.35})MnO_3$ during continuous cooling and heating in zero magnetic fields are shown in Fig. 2. Note that the electrical resistivity of $(Sm_{0.65}Sr_{0.35})MnO_3$ at 4.2 K is $\rho \approx 0.2$ $\Omega \cdot cm$. At 103 K and 107 K on cooling and heating, respectively, the electrical resistance exhibits a very sharp peak of very large magnitude. Note that the electrical resistance of $(Sm_{0.65}Sr_{0.35})MnO_3$ shows thermal hysteresis of about 3 K, typical for a first order phase transition.[7] The amplitude of the peak on cooling is about 140 $\Omega \cdot cm$, indicating the electrical resistance increases by about 750-fold when compared with that at 4.2 K. Similar to earlier reported data for various $(R_{1-x}A_x)MnO_3$ systems, the entire temperature region can be divided into the three regions: low-temperature, high-temperature, and mid-temperature region where the electrical resistance reflects electronic phase transformations. The electrical resistance of $(Sm_{0.65}Sr_{0.35})MnO_3$ behaves metallic-like ($d\rho/dT > 0$) and semiconducting-like ($d\rho/dT < 0$) below and above the observed peak which is typical for this class of materials.[23]

The electrical resistance of $(R_{1-x}A_x)MnO_3$ in the high-temperature region with $d\rho/dT < 0$ was described by various models including hopping of magnetic polarons and a thermally activated process.[2] However, the nature of the electrical conductivity of $(R_{1-x}A_x)MnO_3$ in the high-temperature insulating state still is not fully understood. According to Borges et al.,[22] the electrical resistance of $(Sm_{0.65}Sr_{0.35})MnO_3$ in the temperature range 130-300 K can be fitted by the variable-range hopping model. Our data for $(Sm_{0.65}Sr_{0.35})MnO_3$ in the temperature range between 130 and 240 K can be fitted by thermal activation law, $R \sim exp(\Delta\varepsilon/kT)$, with the energy gap $\Delta\varepsilon = 23$ meV indicating a possible semiconducting state. However, all electronic states of a semiconductor with such a small energy gap should be degenerate at temperatures above ~250 K resulting in metallic-like behavior of the electrical resistance. Hence, the electrical resistance of $(Sm_{0.65}Sr_{0.35})MnO_3$ is determined by the interactions between conduction electrons and Mn localized magnetic moments which changes the mobility rather than the concentration of charge carriers.

In general our data for the electrical resistance of $(Sm_{0.65}Sr_{0.35})MnO_3$ are similar to those reported by Borges et at.[22] However, we can indicate two distinct differences. First, our sample on heating showed a maximum at 107 K while according to Ref. 22 it is observed at 120 K. Second, the observed peak of the electrical resistance of our sample is higher by 20-fold. It is very likely that one of the most important reasons for the observed differences is the high purity of components used in our studies. Furthermore, the process of the sample annealing may make a difference because the sample studied in Ref. 22 was annealed at 1200°C while our sample was annealed twice at 950°C for 21.5 hours and then at 1250°C for 6 days. Among possible reasons for the observed difference in the electrical resistance could be a local deformation of $MnO_6$ octahedra due to the purity of used components or/and annealing temperature, which is critical for orbital degeneracy and occupation of Mn ions.

The inset on Fig. 2 shows temperature dependencies of the electrical resistance of $(Sm_{0.65}Sr_{0.35})MnO_3$ measured on cooling and heating in a 20 kOe magnetic field. In contrast to the data for zero magnetic field (see Fig. 2), both peaks of the electrical resistance are observed at nearly the same temperature of 120 K (as reported in Ref. 22 for zero magnetic field) and they have approximately 5-fold smaller amplitude. Hence, a magnetic field strongly suppresses the insulating state of $(Sm_{0.65}Sr_{0.35})MnO_3$ and eliminates the thermal hysteresis in the vicinity of phase transformation. This is quite different compared to the $Gd_5(Si_xGe_{4-x})$ alloys with a true FOPT where a magnetic field just increases the temperature of phase transformations between low-temperature ferromagnetic and high-temperature paramagnetic phases.[7,13-16]

The temperature dependencies of the electrical resistance of $(Sm_{0.65}Sr_{0.35})MnO_3$ measured on heating in various magnetic fields are presented on Fig. 3. The applied magnetic field decreases the peak and shifts it to a higher temperature. The inset on Fig. 3 shows that the magnitude of the peak of the electrical resistance decreases nearly exponentially with a magnetic field. Similar phenomena were reported, for example, by Schiffer et al.[24] for $(La_{0.75}Ca_{0.25})MnO_3$.

There are a few models (see, for example, Refs. 1,2) used for the explanation of the observed behavior of the electrical resistance in a magnetic field, including the model of the reduction of localization by spin fluctuation scattering with the increasing of the magnetic field. However, it is clear that the suppression of the insulating state in manganese oxides by a magnetic field is the main reason for the observed large magnetoresistance. Our data also indicate that $(Sm_{0.65}Sr_{0.35})MnO_3$ practically does not undergo a temperature induced transition into the insulating state in high magnetic fields, $H > 40$ kOe, and both the low-temperature ferromagnetic and high-temperature paramagnetic phases are metallic. The inset on Fig. 3 shows that the temperature of the peak of the electrical resistance increases nearly linearly with a rate of 0.57 K/kOe. That fact can be one of the evidences that the metal ↔ insulator transition observed in $(Sm_{0.65}Sr_{0.35})MnO_3$ is a first order phase transition because a magnetic field typically shifts the transition towards higher temperatures. Note that in the $Gd_5(Si_xGe_{4-x})$ alloys with true FOPT, a magnetic field increases the temperature of this transition with a very similar rate of 0.5 K/kOe.[7,14,15]

Figure 4 shows the electrical resistance of $(Sm_{0.65}Sr_{0.35})MnO_3$ vs. applied magnetic field at various temperatures ranging between 80 and 134 K.

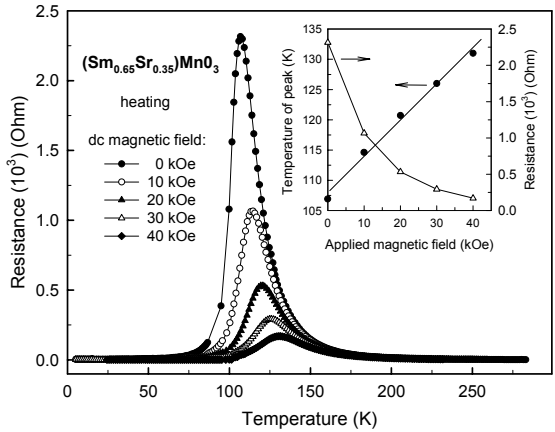

FIG. 3. Temperature dependencies of the electrical resistance of $(Sm_{0.65}Sr_{0.35})MnO_3$ measured on heating in magnetic fields varying from 0 to 40 kOe. The inset shows the dependence of the electrical resistance peak amplitude and its temperature vs. magnetic field.

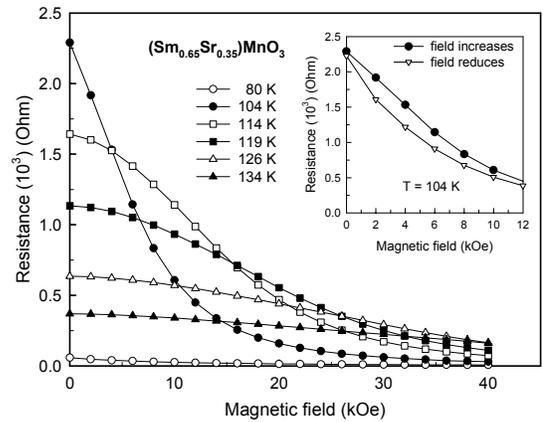

FIG. 4. Magnetic field dependencies of the electrical resistance of $(Sm_{0.65}Sr_{0.35})MnO_3$ measured at temperatures varying from 80 to 134 K. The inset shows the magnetic field hysteresis of the electrical resistance at 104 K.

Below and above these temperatures, respectively, the electrical resistance of $(Sm_{0.65}Sr_{0.35})MnO_3$ weakly depends on the magnetic field. However in the vicinity of the phase transition, at 104 K, the electrical resistance decreases drastically with magnetic field showing the largest rate of ~0.2 Ω/Oe when the magnetic field changes between zero and approximately 8 kOe. The observed decreasing of the electrical resistance is the result of the suppression of the insulating state in $(Sm_{0.65}Sr_{0.35})MnO_3$ (see also Fig. 3). The magnetoresistance at 104 K calculated as *[R(0)-R(H)]/R(H)*, typically used for materials with CMR, is ~8000 % in the 40 kOe magnetic field (it should be noted that classical definition of the magnetoresistance is *[R(H)-R(0)]/R(0)* and negative magnetoresistance has a limit of 100 %). Below and above the electronic phase transition, the magnetoresistance of $(Sm_{0.65}Sr_{0.35})MnO_3$ is negative and small.

The inset of Fig. 4 shows the electrical resistance of $(Sm_{0.65}Sr_{0.35})MnO_3$ at 104 K when the magnetic field was isothermally increased from zero to 50 kOe and then reduced to zero again. One can see the hysteresis is largest between 2 and 6 kOe, reflecting that the electrical resistance here has the highest sensitivity to a magnetic field (see curve for 104 K on Fig. 4). Hence, both the temperature and magnetic field dependencies of the electrical resistance of $(Sm_{0.65}Sr_{0.35})MnO_3$ show smooth transformations between metallic and insulating states with hysteretic character. As was mentioned above, the magnetoresistance of materials with FOPT reflects, in principle, the difference between the electrical resistance of the initial and transformed phases. Dramatic differences between the magnitude of the magnetoresistance of $(Sm_{0.65}Sr_{0.35})MnO_3$ and $Gd_5(Si_2Ge_2)$ take place because the electronic transformation in these materials is metal ↔ insulator and metal ↔ metal, respectively.

### B. Magnetic properties in the dc and ac magnetic field

Figure 5 shows the temperature dependencies of the magnetization of $(Sm_{0.65}Sr_{0.35})MnO_3$ measured on cooling and heating in a 10 kOe magnetic field. The behavior of the magnetization is typical, in general, for materials with ferromagnetic ↔ paramagnetic FOPT induced by temperature. Thermal hysteresis between approximately 80 and 140 K is similar to that observed for the electrical resistance (see also Fig. 2) and it could be additional evidence that the magnetic transition observed in $(Sm_{0.65}Sr_{0.35})MnO_3$ at a temperature around 100 K is a first order transition. The inset of Fig. 5 shows a temperature dependence of the inverse dc magnetic susceptibility of $(Sm_{0.65}Sr_{0.35})MnO_3$. Above 150 K this dependence can be fitted by the Curie-Weiss law $\chi_{dc}(T)=Np^2_{eff}/3k(T-\theta_p)$ with the paramagnetic Curie temperature $\theta_p \approx 100$ K and the effective magnetic moment $p_{eff} = 8.2$ $\mu_B$/f.u. This is larger compared to that expected for the Mn ions; the theoretical values for the $Mn^{4+}$ and $Mn^{2+}$ ions are ~3.8 to and 5.9 $\mu_B$, respectively. The magnetization of $(Sm_{0.65}Sr_{0.35})MnO_3$, measured at temperatures varying between 39 and 157 K, is presented in Fig. 6. At 39 K, the magnetization shows ferromagnetic state with a magnetic field at saturation of ~8 kOe, coercive field $H_c = 0.3$ kOe, and remanent magnetization $M_r = 10$ emu/g. The effective magnetic moment of $(Sm_{0.65}Sr_{0.35})MnO_3$ in the saturated state, extrapolated to zero-magnetic field, is 3.4 $\mu_B$, which is smaller compared to the theoretical value indicating that at low temperatures some fraction of the Mn ions may be coupled antiferromagnetically. Similar magnetization is observed below 39 K, while between 79 K and 132 K the magnetization exhibits a non-linear increase with a magnetic field. Also, hysteresis with a maximal width at ~100 K is observed in the vicinity of the magnetic and electronic phase transformations. At 157 K, in the paramagnetic region, the magnetization shows the change of a slope at ~25 kOe. Such features were already observed for $(Sm_{0.65}Sr_{0.35})MnO_3$ at 160 K and has been explained by a changing of size of magnetic clusters,[22] suggesting that some short-range exchange interactions exist even in the paramagnetic state.

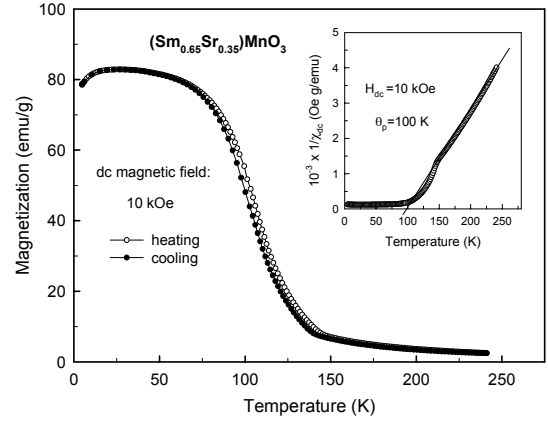

FIG. 5. Temperature dependencies of the magnetization of $(Sm_{0.65}Sr_{0.35})MnO_3$ measured on cooling and heating in the 10 kOe magnetic field. The inset shows temperature dependence of the inverse dc magnetic susceptibility of $(Sm_{0.65}Sr_{0.35})MnO_3$.

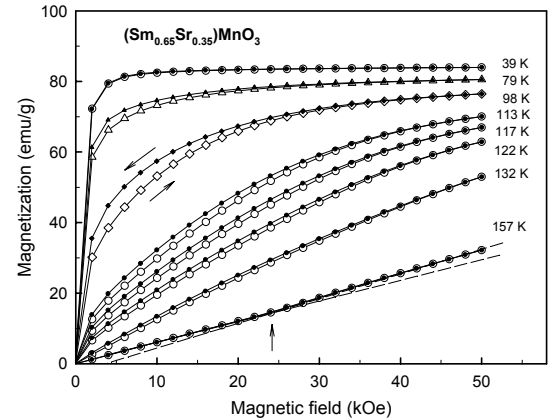

FIG. 6. Isothermal magnetization of $(Sm_{0.65}Sr_{0.35})MnO_3$ at various temperatures.

In should be noted here that similar hysteretic behavior of the magnetization and electrical resistivity measured on cooling and heating was also observed[25,26] for $Pr_xSr_{1-x}MnO_3$, $x = 0.50$ and $0.51$, where the high temperature phase in contrast to $Sm_{0.35}Sr_{0.65}MnO_3$ is antiferromagnetic. $^{55}Mn$ NMR spectra of $Pr_xSr_{1-x}MnO_3$[25,26] and some other manganese oxides[27] show the two-phase character of these manganites clearly indicating an intrinsically inhomogeneous magnetic state which typically is observed in the vicinity of a first order phase transition.

Figure 7 shows temperature dependencies of the real ($\chi'_{ac}$) and imaginary ($\chi''_{ac}$) components of the ac magnetic susceptibility of $(Sm_{0.65}Sr_{0.35})MnO_3$ measured in 12.5 and 25 Oe ac magnetic fields, at 125 Hz. Note that the ac magnetic susceptibility allows the measurement of the magnetic parameters of a material without strong magnetization as takes place when measurements are conducted in a dc magnetic field. The temperature dependence of $\chi'_{ac}$ of $(Sm_{0.65}Sr_{0.35})MnO_3$ above 130 K is similar to the magnetic susceptibility measured in the dc magnetic field and can be fitted by the Curie-Weiss law with the paramagnetic Curie temperature $\theta_p \approx 100$ K and effective magnetic moment $p_{eff} = 7.9$ $\mu_B$/f.u., which are very close to those obtained from the dc magnetic susceptibility. Hence, the high temperature phase has a paramagnetic state in a small magnetic field (see the inset on Fig. 7a). A positive paramagnetic Curie temperature typically is associated with a ferromagnetic low-temperature state and, therefore, both the dc and ac magnetic susceptibility support data[22] that the magnetic phase transformation observed in $(Sm_{0.65}Sr_{0.35})MnO_3$ is between low-temperature ferro-magnetic and high-temperature paramagnetic phases.

The increase in $\chi'_{ac}$ just below 130 K shows the development of strong exchange interactions and formation of long-range ferromagnetic order. Note that $\chi'_{ac}$ changes nearly linearly in the temperature range between approximately 70 and 100 K. Furthermore, $\chi'_{ac}$ of $(Sm_{0.65}Sr_{0.35})MnO_3$ measured in both the 12.5 and 25 Oe ac magnetic fields shows some feature at ~75 K and a sharp peak at ~30 K and there is no large difference between $\chi'_{ac}$ measured in these fields.

Our data for $\chi'_{ac}$ are, in general, similar to that reported by Borges *et al.*[22] including the observed features at 30 and 70 K. Usually a peak in the ac magnetic susceptibility in non-soft ferromagnets is observed around Curie temperature, $T_C$, as a result of increasing $\chi'_{ac}$ with lowering temperature to $T_C$ (determined by common behavior of the magnetization of disordered paramagnetic ions) and then decreasing below $T_C$ (determined by the behavior of the magnetization of ferromagnetic domains and domain walls in low ac magnetic field). Such a decrease is observed because a low ac magnetic field is unable to shift the domain walls. Note that the ac magnetic susceptibility of soft ferromagnets below $T_C$ is practically temperature independent (see, for example, data for ferromagnetic $GdAl_2$, $DyAl_2$, and $ErAl_2$ intermetalics[28]). Hence, in case of $(Sm_{0.65}Sr_{0.35})MnO_3$ the observed peak of $\chi'_{ac}$ means that below ~30 K, in the ferromagnetic region, the coercivity increases very sharply and it may be a reason why the material looks, for a low ac magnetic field, like a frozen spin system. Note that the magnetization of $(Sm_{0.65}Sr_{0.35})MnO_3$ measured in the 10 kOe dc magnetic field (see Fig. 5) also shows a small decrease below ~10 K.

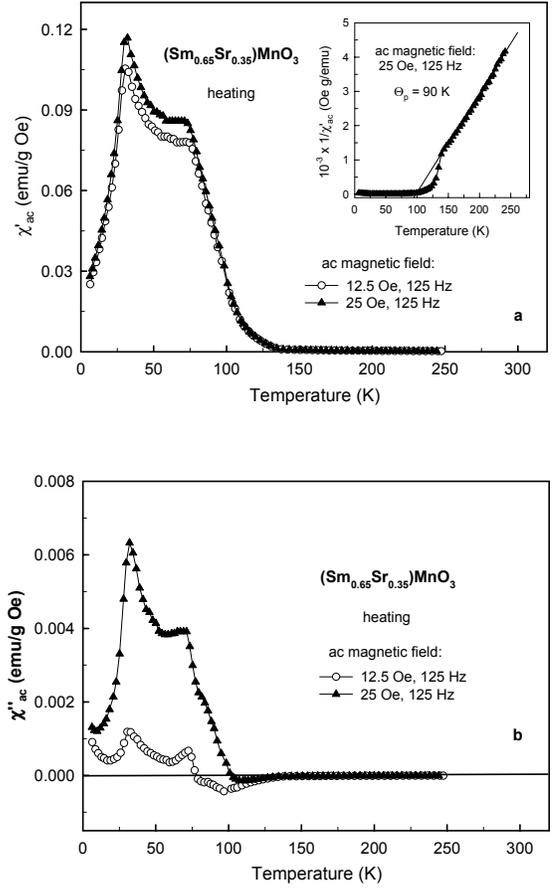

FIG. 7. Temperature dependencies of the (a) real and (b) imaginary components of the ac magnetic susceptibility of $(Sm_{0.65}Sr_{0.35})MnO_3$ measured on heating in the 12.5 and 25 Oe, both 125 Hz ac magnetic field. The inset on Fig. 7a shows temperature dependence of the inverse real component of the ac magnetic susceptibility of $(Sm_{0.65}Sr_{0.35})MnO_3$.

The large value of the magnetic moment in the paramagnetic state reported by Borges et al.[22] was explained by the existence of clusters containing ~30 Mn ions while our data show smaller clusters. However, the development of ferromagnetic order in the layered crystal structure of $(Sm_{0.65}Sr_{0.35})MnO_3$ below 130 K includes the process of the transformation of the short-range order within nearest Mn ions into the long-range magnetic order propagated through the material. During this transformation both the thermal and magnetic hysteresis in the electrical resistance (see Fig. 2 and inset of Fig. 4) and magnetization (Fig. 5 and 6) are observed. When the magnetic hysteresis is close to zero, i.e., in the magnetic field above ~20 kOe (Fig. 6), the thermal hysteresis is also close to zero (see temperature dependencies of the electrical resistance in 20 kOe on the inset of Fig. 2). Hence both the electronic and magnetic transformations in $(Sm_{0.65}Sr_{0.35})MnO_3$ are connected to each other but the applied magnetic field suppresses only the electronic transition creating a metallic state in both the ferromagnetic and paramagnetic states.

Temperature dependencies of the imaginary component of the ac magnetic susceptibility of $(Sm_{0.65}Sr_{0.35})MnO_3$ in 12.5 and 25 Oe ac magnetic fields are similar to $\chi'_{ac}$ showing the two-peak structure (Fig. 7b). In general, the magnitude of $\chi''_{ac}$ in ferromagnets reflects the absorption of energy by the domain walls during their excitation by the ac magnetic field.[29,30] The increase of $\chi''_{ac}$, i.e. energy losses, with the magnitude of the ac magnetic field observed for $(Sm_{0.65}Sr_{0.35})MnO_3$ agrees well with that model. Note that $\chi''_{ac}$ measured in 25 Oe sharply increases below ~105 K, indicating that the ferromagnetic phase and related domain structure occur around this temperature.

## C. Magnetic and electronic phase transformations

Now let's compare the characteristics of phase transformations observed for $(Sm_{0.65}Sr_{0.35})MnO_3$ with those of intermetallic alloys with a true FOPT, for example, Si-rich $Gd_5(Si_xGe_{4-x})$ alloys.[7,14,15] Both materials have similar high-temperature paramagnetic and low-temperature ferromagnetic states, and the transition between these states can be induced by temperature. Note also that the basic ternary manganese oxide $LaMnO_3$ has a magnetic structure where spins couple ferromagnetically in the *xy* layers but antiferromagnetically along the *z*-axis, which is very similar to that suggested for $Gd_5Ge_4$.[17] However, the magnetic transformation in the intermetallic alloys is definitely first-order, with large thermal (6 K) and magnetic field (8 kOe) hysteresis.[15] These are some classical features of FOPT, which also were observed in $Pr_{0.6}Ca_{0.4}Mn_{0.96}Ga_{0.04}O_3$[12] but not in $(Sm_{0.65}Sr_{0.35})MnO_3$. The latter cannot be transformed from the high-temperature paramagnetic state into the ferromagnetic state by a magnetic field as is possible for the $Gd_5(Si_xGe_{4-x})$ alloys.

The magnetoresistance of these intermetallic alloys mainly reflects the difference between the initial phase and transformed phase[7,15] but both are metallic-like. The magnetoresistance of $(Sm_{0.65}Sr_{0.35})MnO_3$ is a result of the change of the electrical resistance due the transformation of a high-temperature insulating-like state into a metallic state induced by a magnetic field. However, in some manganese oxides, both the temperature and magnetic field induced transformations are very similar to those observed for intermetallic alloys mentioned above. Hardy et al.[12] showed that the phase transformation observed for $Pr_{0.6}Ca_{0.4}Mn_{0.96}Ga_{0.04}O_3$ has the same martensitic character as that for the $Gd_5Ge_4$ compound, which represents the Ge-end of the $Gd_5(Si_xGe_{4-x})$ series. Both materials exhibit FOPT, antiferromagnet ↔ ferromagnet, which can be induced by temperature or/and magnetic field. Because antiferromagnetic and ferromagnetic phases can co-exist in these materials under certain conditions, despite the quite different compositions of $Pr_{0.6}Ca_{0.4}Mn_{0.96}Ga_{0.04}O_3$ and $Gd_5Ge_4$, they represent similar magnetic phase-separated systems. The different character of the magnetic phase transformations observed in $(Sm_{0.65}Sr_{0.35})MnO_3$ indicates that a martensitic transformation in this oxide is not very likely.

## IY. CONCLUSIONS

A $(Sm_{0.65}Sr_{0.35})MnO_3$ sample prepared from high purity components has a monoclinically distorted crystal structure when compared to a cubic perovskite-type sample reported on earlier. Below ~100 K, $(Sm_{0.65}Sr_{0.35})MnO_3$ has a ferromagnetic state and during heating in zero magnetic field, it transforms into the paramagnetic state above 100 K. Simultaneously with the magnetic phase transformation, a metal → insulator electronic transition is observed, with a change of the electrical resistance from ~3 Ω to ~2000 Ω between 4.2 K and 106 K. The character of the transformations is similar to that observed for metallic materials with a first order phase transformation. The magnetic field suppresses the metal ↔ insulator transition of $(Sm_{0.65}Sr_{0.35})MnO_3$ and in H > 40 kOe it has the metallic state in the wide temperature range.

The magnetoresistance of $(Sm_{0.65}Sr_{0.35})MnO_3$ in both the metallic and insulating states is negative and small, a few %. In the vicinity of the metal ↔ insulator transition it also is negative but has "colossal" character reaching ~8000 % in a 40 kOe magnetic field. The real and imaginary components of the ac magnetic susceptibility of $(Sm_{0.65}Sr_{0.35})MnO_3$ confirms the transition between low-temperature ferromagnetic and high-temperature paramagnetic states.

So, some magnetic and magnetoelectronic properties of $(Sm_{0.65}Sr_{0.35})MnO_3$ are different compared to those reported earlier for the same composition, which is probably due to differences in the purity of used components and/or different process of annealing which could effect a local deformation of $MnO_6$ octahedra. There are a few similar features observed for $(Sm_{0.65}Sr_{0.35})MnO_3$ and $Gd_5(Si_xGe_{4-x})$ materials but the character of the phase transformation is, in general, different.

ACKNOWLEDGMENT

E.M.L. would like express a gratitude to Prof. K.A. Gschneidner, Jr. for his support and interest to this work, Prof. V.K. Pecharsky for his support and help with the analysis of x-ray powder diffraction data, and Dr. V.P. Balema for providing of the sample. Authors are grateful to Dr. S.L. Bud'ko for helpful discussions and valuable comments. This manuscript has been authored by Iowa State University of Science and Technology under Contact No.W-7405-ENG-82 with the U.S. Department of Energy, and partially supported by the Office of Basic Energy Sciences, Materials Science Division.